# Brillouin scattering and molecular dynamics study of the elastic properties of Pb(Mg$_{1/3}$Nb$_{2/3}$)O$_3$

Muhtar Ahart[1*], Aravind Asthagiri[1,3], Zuo-Guang Ye[2], Przemyslaw Dera[1],

Ho-Kwang Mao[1], R. E. Cohen[1], and Russell J. Hemley[1]

[1]Geophysical Laboratory, Carnegie Institution of Washington, 5251 Broad Branch Rd.,

NW, Washington DC 20015

[2]Department of Chemistry, Simon Fraser University, Burnaby, B. C., V5A 1S6, Canada

[3]Department of Chemical Engineering, University of Florida, Gainesville, FL 32611

We report the complete set of elastic constants and the bulk modulus for single crystal Pb(Mg$_{1/3}$Nb$_{2/3}$)O$_3$ (PMN) at room temperature obtained from Brillouin spectroscopy and Molecular Dynamics (MD) simulations. The bulk modulus from Brillouin is found to be 103 GPa, in good agreement with earlier x-ray studies. We also derived the refractive index along all principal axes and found PMN to be optically isotropic, with a refractive index value of 2.52 ± 0.02. PMN shows mechanical anisotropy with $A$=1.7. The MD simulations of PMN using the random site model overestimate the elastic constants by 20-50 GPa and the bulk modulus is 148 GPa, but the mechanical anisotropy matches the Brillouin results of $A = 1.7$. We also determined the elastic constants for various models of PMN and we find variation in the elastic constants based on chemical ordering.

**PACSnumbers:** 77.90.+k, 77.65.-j, 78.35.+c

[*] Corresponding author: maihaiti@ciw.edu

**I. Introduction**

There has been a great deal of interest in a new class of relaxor ferroelectrics, such as $(1-x)Pb(Mg_{1/3}Nb_{2/3})O_3$-$x$PbTiO$_3$ (PMN-$x$PT) and $(1-x)Pb(Zn_{1/3}Nb_{2/3})O_3$-$x$PbTiO$_3$ (PZN-$x$PT). This interest has been fueled by the superior electromechanical properties[1] of these materials compared to the conventional PbZr$_x$Ti$_{(1-x)}$O$_3$ ceramics that have dominated piezoelectric applications for more than 40 years. While extensive theoretical and experimental studies[1-5] have greatly advanced our understanding of relaxors, many properties of relaxors still remain unclear.[6] The difficulties stem from the complexity of these materials, which have a high degree of compositional, chemical and structural disorder.

Pb(Mg$_{1/3}$Nb$_{2/3}$)O$_3$ (PMN) is a prototypical relaxor and an end-member of the technologically important relaxor ferroelectric solid solution PMN-PT. PMN has a perovskite structure with a cubic *m3m* symmetry,[7] with Mg$^{2+}$ and Nb$^{5+}$ ions disordered on the B-sites. Measurements of the dielectric constants show that the PMN crystals do not undergo a sharp transition to a ferroelectric phase. Instead, the dielectric constant exhibits a broad maximum at $T_{max}$~ 270 K.[8] Below T~620 K, the dielectric constant deviates from the Curie-Weiss law and exhibits strong frequency dispersion below 350 K.[8] Relaxors also show x-ray and neutron diffuse scattering[9,10] along with broad Raman spectra.[11,12] Brillouin scattering was previously used to investigate the temperature dependences of longitudinal (L-) and transverse (T-) modes of PMN in backscattering geometry,[13,14] which showed a minimum in temperature dependence of sound velocity near T ≈ 240 K for L- and T-modes. The phase diagram of the solid solution PMN−$x$PT contains a rhombohedral phase for 0< x < 33%; an intermediate phase associated with the morphotropic boundary (MPB) for 33%< x < 38%; and a tetragonal phase for x > 38% at ambient pressure and room temperature.[15, 16] Furthermore, one end-member (PMN) is a typical relaxor, while the other end-member (PT) is a conventional ferroelectric material, and the composition with the largest value of electromechanical coupling coefficient is located near the MPB.

Knowledge of elastic, piezoelectric, and dielectric constants is fundamental for describing the electromechanical response of a material to applied strains and fields, so a complete set of single crystal tensor properties as a function of composition is desirable. Recently, a complete set of elastic and piezoelectric constants has been reported for multidomain and single domain PZN-$x$PT and PMN-$x$PT at several different compositions.[17-27] Elastic properties can be determined by using different methods such as ultrasonic resonance or Brillouin scattering techniques, depending on the availability of single crystals of appropriate size. Of these techniques Brillouin scattering offers the possibility to probe very small regions (tens of microns) in the sample. Earlier Brillouin studies of PMN were limited to backscattering geometries and therefore all the elastic constants could not be derived. In this paper, we study the elastic properties of PMN single crystal by using the micro-Brillouin scattering technique in a symmetric scattering geometry. We have also performed molecular dynamics (MD) simulations using a recently developed shell model potential[28] to extract the elastic constants of PMN. Validated MD simulations for PMN-$x$PT would be valuable since we can extract many material properties, including the electromechanical properties, with far greater ease from simulations than experiment and furthermore can determine the impact of different proposed models for PMN. The rest of the paper is organized as follows: Section II and III outline the experimental and simulation details. In section IV, we discuss the elastic constants obtained for PMN from Brillouin spectroscopy and MD simulations. We also compare the elastic properties obtained for pure PMN with literature results for PMN-$x$PT at several compositions and discuss correlations that can be gleaned from the current database of electromechanical constants.

**II. Experimental Details**

Single crystals of PMN with cubic morphology were grown by the high-temperature solution technique.[29] We have confirmed by x-ray analysis that the material has point group symmetry $m3m$. A sample polished normal to the ab plane with the thickness of about 300 μm was mounted on a thin glass plate with silicone grease. Neither birefringence nor domains were visible by polarized microscope, which is consistent with the cubic symmetry. We also examined the sample after the Brillouin measurements and observed no changes in linear optical properties.

A full description of the Brillouin scattering apparatus used in this study can be found elsewhere.[30] Experimental details specific to the data collection for PMN are similar to those of a previous PbTiO$_3$ study.[31] Light from a single mode Ar-ion laser (λ=514.5 nm) was used as the excitation source with the average power less than 100 mW. The sample was placed symmetrically with respect to the incoming and collected light such that the difference vector of the two beams was in the plane of the sample. A diagram of the ($\theta$=80°) symmetric scattering geometry is shown in Fig. 1. By rotating the sample around the [001] direction, the scattering vector is rotated in the ab plane of the sample. The scattered light was analyzed by a 3+3 tandem Fabry-Perot interferometer, detected by a photon counting photomultiplier and output to a multichannel scalar. Spectra were taken every 10° over a full rotation of the sample and collected for an average of two hours at each single orientation.

**III. Modeling Details**

An accurate shell model potential was recently developed for PMN-$x$PT by fitting to first-principles data of PbTiO$_3$ and ordered supercells of PMN. We briefly discuss the potential, but refer the reader to Ref. 28 for more details of the development of the potential. The shell model represents each atom by a positively charged core and negatively charged shell and the core-shell displacement induces atomic polarizability. The core and shell are connected by an anharmonic spring, $V(w) = \frac{1}{2}c_2 w^2 + \frac{1}{24}c_4 w^4$, where w is the core-shell displacement. We have found it necessary to add an exponential term, $D(w-w_0)^2$ if $w \geq w_o$ where $w_o = 0.2\,\text{Å}$ and $D$ is set to 10000 eV, to the core-shell coupling to prevent the shell from drifting off the core and ensure the potential stability. There are Coulombic interactions between all cores and shells except the core and shell of the same atom. There are short-range interactions between the shells of the atoms and we use the Rydberg potential form, $V(r) = (A+Br)e^{(-r/\rho)}$, for the A-O, B-O, and O-O shells.

In section IV, we present elastic constants obtained from MD simulations using our shell model potential for various models of PMN. The potential has already been shown to qualitatively reproduce the compositional phase diagram for PMN-$x$PT.[28] In particular, the transition between the rhombohedral low PT composition phase and the tetragonal high PT composition phase occurs through a monoclinic-type phase. Furthermore, this transition occurs near 33% as seen experimentally, but the phase is broader in range than experimentally found and does not become fully tetragonal until it exceeds 50% PT concentration. Pure PMN was found to be macroscopically cubic similar to experiment and the temperature and pressure phase diagram for PbTiO$_3$ derived from the potential matches well with experiment.[28] All MD simulations reported in this paper are carried out using DL-POLY[32] and unless explicitly stated are performed in an NσT ensemble at 300 K and zero stress. A supercell size of 6×6×6 (1080 atoms) was used with periodic boundary conditions. A mass of 7 a.u. was assigned to the Pb, Nb, and Ti shells and 2 a.u. to the O and Mg shells. The time step was set to 0.2 fs. Unless otherwise noted a single MD simulation run consists of 500000 time steps with data collected after 100000 time steps. We have used these relatively long simulation runs to ensure that the elastic constants extracted from the MD runs are converged.

The elastic constants can be calculated from MD simulations using the thermal stress or strain fluctuations. The two methods have relative advantages and disadvantages. The stress fluctuation method[33-35] involves additional potential derivative terms that must be calculated in the MD simulations, while the strain fluctuation method[36] converges slowly. We have used a method proposed by Gusev and co-workers based on the correlation between the stress and strain fluctuations.[37] This method is more efficient than the strain fluctuation method and only requires tracking the instantaneous stress and strain tensors during the MD simulation. The formula for the elastic constants is given by

$$C_{iklm} = \langle \varepsilon_{ik} \sigma_{nj} \rangle \langle \varepsilon_{nj} \varepsilon_{lm} \rangle^{-1}, \qquad (1)$$

where $\sigma_{nj}$ are the components of the instantaneous stress tensor and $\varepsilon_{ik}$ is the instantaneous strain tensor and < > indicates an average over time. The strain tensor is given by

$$\varepsilon_{ik} = \frac{1}{2}(h_{nl}\langle h \rangle^{-1}_{lk} h_{np} \langle h \rangle^{-1}_{pi} - \delta_{ik}), \qquad (2)$$

where $h_{ik}$ are the components of the scaling matrix, $h$=(**a**,**b**,**c**), made up of the vectors of the simulation cell. The stress term ($\sigma_{nj}$) that appears in Eq. (1) involves assuming that the differences between the average and instantaneous scaling matrix is small, a valid assumption for the low temperatures examined in this paper.[37]

The shell model potential was not derived by fitting to any electromechanical constants; therefore it is important to estimate the error in the elastic constants. To evaluate this error we initially examine PbTiO$_3$ at 300 K, which is a far simpler and better characterized experimentally than PMN. We also evaluate the accuracy of the stress-strain fluctuation method, by comparing elastic constants obtained via Eq. (1) versus the direct method. In the direct method small strains are applied on the zero stress structure and the average stress is measured from a NVT MD simulation. For the direct method since we are interested in only the average stress tensor not the fluctuations we can use much shorter simulations consisting of 200000 time steps. We apply strains ranging from -0.5 % to 0.5% and the elastic constants are determined from the slope of the stress versus strain line. The direct method is straightforward and does not rely on fluctuation formulae, but requires a multitude of MD simulations for each elastic constant. Table 1 shows the elastic constants obtained from our MD simulations along with experimental and Density Functional Theory (DFT) values from literature. The stress-strain fluctuation method gives excellent agreement with the direct method. Furthermore, we see good agreement between experiment and our MD simulations for all the elastic constants except for $C_{66}$, which is overestimated by ~ 30 GPa. For $C_{33}$ and $C_{13}$ there is variability in the experimental values because a unique solution could not be derived from the Brillouin data.[38] For some elastic constants the DFT results are actually worse than our MD results when compared to experimental values. The error in DFT has been well documented and can be mainly attributed to the exchange-correlation functional. The potential was derived from fitting to DFT data, but incorporates both PbTiO$_3$ and PMN input data and furthermore did not include any elastic constants as input. Therefore, the potential will deviate from the PbTiO$_3$ DFT data and in the case of pure PbTiO$_3$ this deviation moves the results closer to experiment. We will discuss the elastic constants obtained from MD simulations for PMN in the next section.

**IV. Results and Discussion**

A set of typical Brillouin spectra is shown in Fig. 2. There are two pairs of peaks in each spectrum and an additional Rayleigh peak due to elastic scattering at zero frequency. One of the two pairs corresponds to the longitudinal acoustic mode (L-mode) and the other pair corresponds to the transverse acoustic modes (T-mode).

Brillouin scattering due to the inelastic interaction of light with acoustic phonon in the crystal results in a frequency shift $\Delta v$ of the incident light. This is related to the velocity of the acoustic phonon $\upsilon$ propagating with the wave vector $q$ by [39]

$$\upsilon = \frac{\Delta v \lambda}{(n_i^2 + n_s^2 - 2n_i n_s \cos\theta)^{1/2}}, \qquad (3)$$

where $n_i$ and $n_s$ are the refractive indices for the incident and scattered light, and $\theta$ is the scattering angle. In this study, the sample is placed symmetrically with respect to the incoming and collected light. For this particular geometry, the frequency shift of the incident light is independent of the refractive index $n$ of the sample (detailed derivation can be found in Ref. 30). The Brillouin shifts $\Delta v_\theta$ are given by

$$\Delta v_\theta = (2\upsilon/\lambda)\sin(\theta/2), \qquad (4)$$

where $v$ is sound velocity and $\lambda$ is the incident laser light wavelength.

The velocity of a plane acoustic wave propagating in a direction $q$ is related to the elastic properties of the crystal via the Christoffel equation:

$$|\Gamma_{ik} - \delta_{ik}\rho v^2| = 0 \qquad (5)$$

with

$$\Gamma_{ik} = \{C_{ijkl}^E + [(e_{mij}q_m)(e_{nkl}q_n)/(\varepsilon_{rs}^s q_r q_s)]q_m q_n\}q_j q_l, \qquad (6)$$

where $\delta_{ik}$ is the Kronecker delta with $\delta_{ik}=1$ for $i=k$ and $\delta_{ik}=0$ for $i \neq k$. The $\rho=8.12$ g/cm$^3$ is the density of PMN. $C_{ijkl}^E$, $e_{mij}$ and $\varepsilon_{rs}^s$ are components of the elastic, piezoelectric stress, and dielectric permittivity tensors, respectively. The superscript $E$ indicates the condition of constant electric field, while the superscript $S$ indicates the condition of constant strain, or clamped permittivity. The $q_i$'s are the directional cosine of the acoustic wave. There are four independent non-vanishing constants for cubic (*m3m*) symmetry, which in Voigt matrix notation are $C_{11}^E$, $C_{12}^E$, $C_{44}^E$, and $\varepsilon_{11}^S$. The dielectric constant $\varepsilon_{11}^S$ is not constrained by our measurements because there is no piezoelectric response in PMN and thus no contribution from the second part of Eq. (4) to the Brillouin shifts. Equation (5) is a cubic equation with three different roots, $\rho v_i^2$ i=1, 2, 3, associated with one longitudinal and two transverse waves propagating in a given direction. For instance, the three roots of the Eq. (5) in the a-b plane of PMN crystals can be expressed as follows:

$$\rho v_L^2 = \frac{C_{11} + C_{44} + \sqrt{(C_{11} - C_{44})^2 \cos^2 2\psi + (C_{12} + C_{44})^2 \sin^2 2\psi}}{2}, \qquad (7)$$

$$\rho v_{T1}^2 = \frac{C_{11} + C_{44} - \sqrt{(C_{11} - C_{44})^2 \cos^2 2\psi + (C_{12} + C_{44})^2 \sin^2 2\psi}}{2}, \qquad (8)$$

$$\rho v_{T2}^2 = C_{44}, \qquad (9)$$

where $\psi$ is the angle of between phonon propagating directions and the a-axis in the a-b plane; the $v_L$, $v_{T1}$, and $v_{T2}$ are the sound velocities of the one longitudinal and two transverse modes. The measured Brillouin shifts were inverted for a best-fit set of elastic constants by minimizing the sum of squares of residuals between the measured Brillouin shifts and those calculated from a trial set of constants using Eqs. 8 and 9 by the method described in Ref. 27. The nonlinear minimization procedure (described in detail in Ref. 27) was used in this process. The comparison of measured Brillouin shifts with those generated from our best-fit parameters is shown in Fig. 3. The fitted curves are in good agreement with the experimental data. Thus, the results indirectly confirm that PMN is cubic on the mesoscopic scale.

Since we measured the full rotation of the velocity curves in a given plane, there is sufficient data to constrain the acoustic velocity in any direction within the plane. Therefore, we can derive all of the elastic constants in contrast to earlier experimental studies of PMN, and we report the parameter set obtained from our best fit of the data in Table II, along with independent values reported in the literature[9, 13] for PMN single crystals. The uncertainties in our reported parameters are given as ±1σ based on the root mean-square-deviation of the measured Brillouin shifts. Our values for $C_{11}$ and $C_{44}$ are in a good general agreement with the values found in literature.[9, 13]

In order to investigate the optical anisotropy, the Brillouin shifts were also measured in the backscattering geometry and the results are shown in Fig. 4. In this case, the phonon propagates along the *a*, *b* or *c* axis, respectively. Since in this particular geometry the Brillouin shifts relate to the refractive index, the Brillouin shifts from the combination of *θ* angle scattering and backscattering will allow us to determine the refractive index along the *a*, *b*, and *c* axis. The results are $n_a = n_b = n_c = \frac{\Delta \nu_{180°}}{\Delta \nu_\theta} \sin(\theta/2) = 2.52 \pm 0.02$. The refractive index value obtained is comparable with previous results n ~ 2.545.[40] Within the error of Brillouin experiments, the crystal is optically isotropic.

We now turn to evaluating the elastic constants for PMN from MD simulations. A central issue in the MD simulations is the handling of the disorder in the B cation site. There is no long range ordering in PMN, but a degree of short range order is expected but the exact nature is still not clear. The current model that fits the existing experimental data is the existence of a disordered matrix with chemical ordered regions (COR) (see Refs. 5, 6 for recent reviews). The COR is described by the random site model, which can be represented as Pb(B',B'')O$_3$ where B' are Nb atoms and B'' are randomly selected but maintaining a 1:1 ratio of Mg:Nb. For a large system the random site model results in an average cubic and isotropic structure. The experimental PMN system would require very large simulations systems (on the order of 50×50×50 unit cells) and obtaining converged elastic constants is computationally demanding, and we do not tackle it directly in this contribution. Instead we examine the elastic constants for PMN composed entirely of the random site model. A recent effective Hamiltonian study of PSN indicates that the COR is equivalent to the polar nanoregions (PNR),[5] which are central to many of the interesting properties of relaxors and therefore it is important to understand their elastic properties.

Since our simulation cell is relatively small (6×6×6) we need to test that the random assignment in the randomsite model does not influence the elastic properties. Therefore, we have performed MD simulations of five different realizations for randomsite PMN (i.e. we generate five different initial configurations) and the resulting elastic constants are all reported in Table 3. For rndsite1 configuration we calculated the elastic constants via the direct method also to ensure that the fluctuation formula (Eq. 1) is accurate and indeed we find no difference between the two methods. The results in Table 3 confirm that a 6×6×6 is sufficiently large system to represent the randomsite model and there is no dramatic influence due to the particular arrangement of the B" atoms. Furthermore, the obtained elastic constants show cubic symmetry as does the overall structure. A comparison of the elastic constants obtained from MD simulations and the Brillouin results in Table 2 show that the MD elastic constants consistently overestimate the Brillouin numbers. For $C_{11}$ this overestimation is close to 60 GPa, while for $C_{66}$ it is closer to 20 GPa and the bulk modulus is 50 GPa larger in MD simulations. Nevertheless, the relative values of the elastic constants are very close to the experimentally observed values.

There can be several sources for the difference between the elastic constants from MD and Brillouin. Firstly, the interatomic potential introduces errors, but at least for PT these errors were not as substantial as observed for PMN. Secondly, as noted earlier the random site model is not a true representation of experimental PMN. The direct simulation of PMN with disordered matrix and embedded COR is not trivial, but we have looked at various ordered models of PMN and completely random PMN to evaluate the sensitivity of the elastic constants to cation ordering. The elastic constants for various models of PMN are reported in Table 3. The $[001]_{MMN}$ and $[111]_{MMN}$ structures have 1:2 order (1 Mg: 2 Nb) along the [001] and [111] planes. The $[001]_{NCC'}$ structure has three different planes along the [001] direction, an Nb plane followed by two planes of (1 Mg: 1 Nb) in the rocksalt structure. Finally, the random structure is a complete random distribution of the B cations with an overall ratio of 1Mg: 2Nb. The macroscopic polarization for all models of PMN is negligible, but as shown in Table 3 the bulk modulus change based on the structure. The ordered models of PMN do not have cubic symmetry due to the chemical ordering. DFT results of these models indicate that they all have low symmetry ground states (monoclinic or triclinic), and this is reflected in the symmetry of the elastic constants we extract from MD. We have also evaluated the elastic constants for the completely random model for PMN, where the 2:1 ratio of Nb and Mg is maintained but the B sites are filled entirely randomly. The individual realizations of the random model do not show cubic symmetry, but after averaging over five realizations we obtain elastic constants that have an average cubic symmetry. Examining the elastic constants between the various models indicates that chemical ordering does impact the elastic properties of PMN. The elastic constants for the random model are closer to experiment than the randomsite model but still show deviation on the scale of 20-30 GPa. While such error is excellent for an interatomic potential that was not fitted directly to any elastic constant data, further studies with more complex structures (i.e. embedded COR's) will be required before the exact impact of the polar nanoregions can be determined.

We conclude with a comparison of the elastic constants of PMN to the material properties across the PMN-$x$PT composition. The PMN crystal exhibits mechanical anisotropy typical for oxides. The elastic anisotropy is defined as $A = 2C_{44}/(C_{11} - C_{12})$ and $A = 2C_{66}/(C_{11} - C_{12})$ for cubic and tetragonal symmetry, respectively. We obtain a value of $A$=1.7 for PMN; it is comparable with MgO, a typical oxide, for which $A$=1.55 (Ref. 41). Interestingly, the elastic anisotropy is exactly the same from both Brillouin spectroscopy and MD simulations. For comparison, the anisotropies of several PMN-PT solid solutions[42] are listed as follows: the multidomain PMN-30%PT, PMN-33%PT, and PMN-42%PT single crystals (poled along the [001] direction) have the macroscopic tetragonal symmetry of 4mm, their values of elastic anisotropy[42] are $A$=9.4, 11, and 1.78, respectively (Fig. 5). One end member, the classic ferroelectric PbTiO$_3$,[31] has anisotropy $A$=1.32. We plot our obtained elastic anisotropy value for PMN along with similar values for PMN-$x$PT reported in the literature in Fig. 5. The PMN-$x$PT crystals in the rhombohedral side of the phase diagram are rotated to a tetragonal symmetry via an applied field in the [001] direction. It is interesting to note that the PMN-$x$PT solid solutions exhibit typical elastic anisotropic values for both end members, but have very large values for compositions near the MPB on the rhombohedral side. The behaviors of the elastic anisotropies for the PMN-$x$PT system correlate well with their electromechanical properties (Fig. 5).

The origin of the high piezoelectric response in relaxor ferroelectric materials has been the subject of both experimental and theoretical studies. The high piezoelectric response is attributed to the ease of rotating the polarization from the rhombohedral to tetragonal orientation by the application of an electric field in the [001] direction in these materials near the MPB. Experimental studies of the phase diagram of PMN-$x$PT and PZN-$x$PT have found the existence of monoclinic phase close to the MPB. This monoclinic phase separates a rhombohedral PT-poor phase from a tetragonal PT-rich phase of PZN-PT or PMN-PT, therefore instead of being constrained to alignment along a symmetric axis the polarization can easily rotate within a monoclinic[43] plane between the tetragonal and rhombohedral phase upon the application of an electric field. Recently, DFT calculations of PbTiO$_3$ under pressure identified a tetragonal-to-monoclinic-to-rhombohedral-to-cubic phase transition sequence and furthermore a colossal enhancement of the piezoelectric response near the phase transitions.[44] This work shows that the existence of the monoclinic phase is not unique to solid solution ferroelectrics. The only necessary condition is that the high symmetry phases are very close in energy and therefore rotation from one phase to the other can occur through the low-symmetry monoclinic phase. In the DFT work on high pressure PbTiO$_3$ they find that the $C_{44}$ elastic constant goes to zero as the tetragonal to monoclinic phase transition is approached and the $C_{11}$-$C_{12}$ (in cubic axis) goes to zero at the rhombohedral to monoclinic phase transition. This result for PbTiO$_3$ matches the observations of an enhanced mechanical anisotropy on the rhombohedral side near the MPB (i.e. $C_{11}$-$C_{12}$ goes to zero therefore $A$ increases) but not on the tetragonal side (i.e. $C_{44}$ goes to zero so $A$ decreases). Preliminary MD simulations of the elastic constants of randomsite PMN-$x$PT show the same qualitative behavior in elastic constants as observed in pressure-induced PT, including the softening of the $C_{44}$ and $C_{11}$-$C_{12}$ elastic constants. We will report a detailed study of the electromechanical constants for PMN-$x$PT derived from MD simulations in the future. While the above explanation would justify the similar observed behavior for PMN-$x$PT, it assumes there are no contributions from domain interactions. The microstructure of PMN-$x$PT is still under much debate [6], and extrinsic contributions due to domains may also influence the electromechanical behavior for these materials. If the extrinsic contributions are minimal, then the electromechanical constants for a domain-engineered sample (i.e. sample that is poled in the [001]) should be equivalent to rotating the single-domain electromechanical constants from rhombohedral to cubic axis.

The above argument also assumes a similarity in the behavior of the elastic constants between the composition induced phase transitions in PMN-$x$PT and the pressure induced phase transitions in PbTiO$_3$. Cao and co-workers have determined the elastic constants for single-domain rhombohedral PMN-33%PT and after rotation to the cubic axis they find similar elastic constants to those obtained from a multi-domain sample poled in the [001] direction.[17, 19] Recently, the electromechanical constants for single-crystal rhombohedral PZN-4.5%PT was obtained from Brillouin, and simple rotation of these single-domain values did not match the elastic constants obtained from poled multi-domain PZN-4.5%PT using macroscopic probes.[27] This result indicates that at least for PZN-PT that extrinsic contributions due to domain interactions does impact the electromechanical properties. Further studies using Brillouin, which probes a much smaller domain size than macroscopic probes, across the PMN-$x$PT composition may assist in clarifying the role of intrinsic and extrinsic contributions to the electromechanical properties of PMN-$x$PT.

## IV. Conclusions

A complete set of elastic constants of PMN single crystal have been derived from micro-Brillouin scattering spectroscopy. We also calculated the bulk modulus for the system and our measurement is in good agreement with the value obtained by x-ray methods. PMN single crystal is an optically isotropic and mechanically anisotropic relaxor ferroelectric at room temperature. We have also evaluated the elastic constants of PMN from MD simulations using a recently developed shell model potential. The MD simulations of various models of PMN indicate that chemical ordering has an impact on the elastic constants of PMN. The elastic anisotropy of PMN-$x$PT solid solutions shows strong compositional dependences. In particular, the composition near the MPB such as PMN-33%PT exhibits an anisotropy of 11, which is also observed in the recently DFT-evaluated elastic constants for high-pressure PbTiO$_3$ near the MPB[44] indicating that the compositional dependence on mechanical anisotropy is mainly due to intrinsic contributions and not domain-domain interactions.

**Acknowledgments:**

This work was sponsored by the Office of Naval Research under Grants No. N00014-02-1-0506, N00014-97-1-0052 and N00014-99-1-0738. Support also was received from the Carnegie/Department of Energy Alliance Center (CDAC, DE-FC03-03NA00144). We acknowledge the University of Florida High-Performance Computing Center (http://hpc.ufl.edu) for providing computational resources for performing the calculations reported in this paper.

**Figure Captions:**

**Figure 1.** A diagram of the symmetric scattering geometry. The incident light enters the sample with the angle $\theta/2$, and by Snell's law the following equation will hold: $n_0\sin(\theta/2)=n\sin(\varphi/2)$, where $n_0$ is the refractive index of air and n is the refractive index of samples. Therefore the Brillouin shifts will have the following form: $\Delta v_\varphi = \frac{2n\upsilon}{\lambda}\sin(\varphi/2) = \frac{2\upsilon}{\lambda}\sin(\theta/2) = \Delta v_\theta$. The Brillouin shifts can be defined uniquely by the scattering angle between incident and scattering light.

**Figure 2.** A set of typical Brillouin spectra at selected angles. The two different modes (L- and T-modes) are clearly visible. The other T-mode, which corresponds to the elastic constant $C_{44}$, did not appear in the spectrum due to the present set up for the polarization of incident light in scattering experiments.

**Figure 3.** Polar plot of the experimental (squares) and model (lines) Brillouin shifts (GHz) of the longitudinal and transverse modes as a function of angle for the ab plane of the PMN crystal.

**Figure 4.** The backscattering spectra of a PMN single crystal. The L-mode corresponds to the acoustic wave propagate along the a-, or b-, or c-axis, respectively. Rayleigh indicates the elastic scattering. Since the backscattering spectrum includes the information of refractive index, we are able to obtain the refractive index along the a-, or b-, or c-axis.

**Figure 5.** Compositional dependences of the elastic anisotropy (open circles) and the piezoelectric constant (solid circles) for the (1-x)PMN-xPT solid solutions. The elastic anisotropy and the piezoelectric constant data were adopted from Refs. 17, 19, 24, 31, and 42 for the composition x=0.3, 0.33, 0.42, and 1. The solid lines (connections between the points) are guides for the eyes.

**Table Captions:**

**Table I.** Elastic constants for tetragonal PbTiO$_3$ obtained from several different methods. Our MD results are obtained at T = 300 K and zero stress, similar to the experimental results. The DFT results are evaluated at the experiment lattice parameters. A unique solution for the experimental elastic constants could not be derived from the Brillouin scattering data [38]. See text for details on the two methods to extract elastic constants from MD simulations.

**Table II.** Material constants for cubic PMN at room temperature obtained from Brillouin.

**Table III.** Elastic constants obtained from Brillouin compared with MD results for PMN at 300 K. The MD simulations used the random site model to represent the disorder on the B cation in PMN. Elastic constants for five distinct realizations of the random site model are shown.

**Table IV.** Elastic constants for various models of PMN obtained from MD simulations at T= 300K and zero pressure. For models that incorporate disorder (i.e. random site and random) the results are from five different realizations.

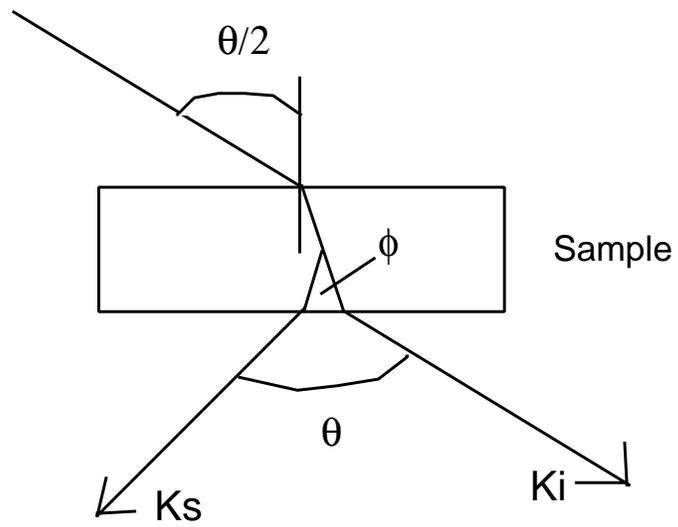

Fig. 1.

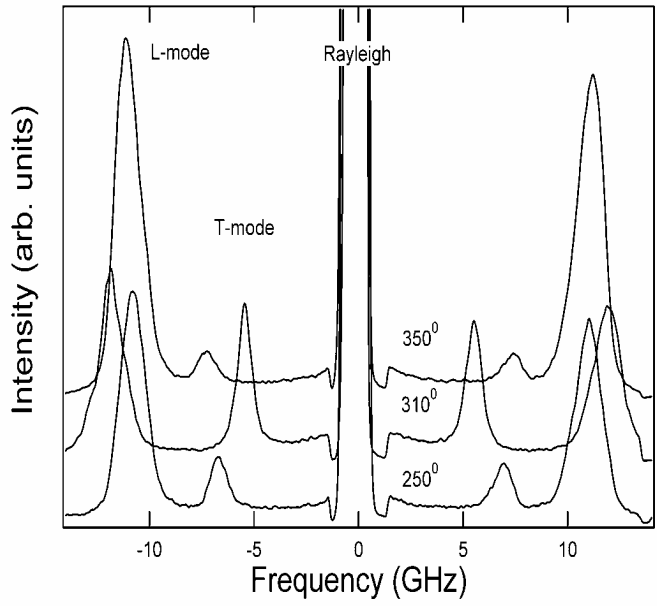

Fig. 2

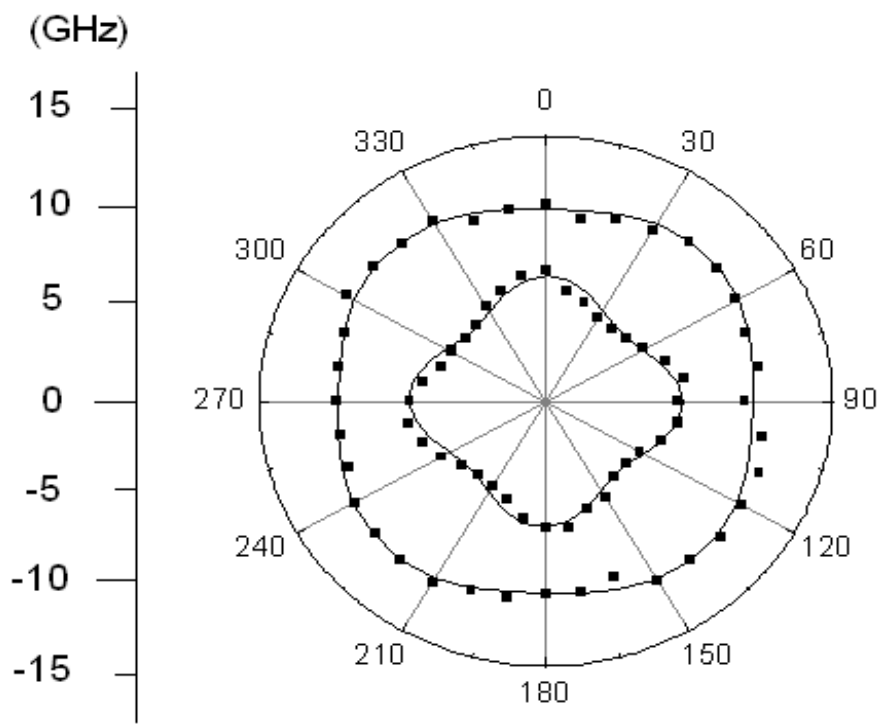

Fig. 3.

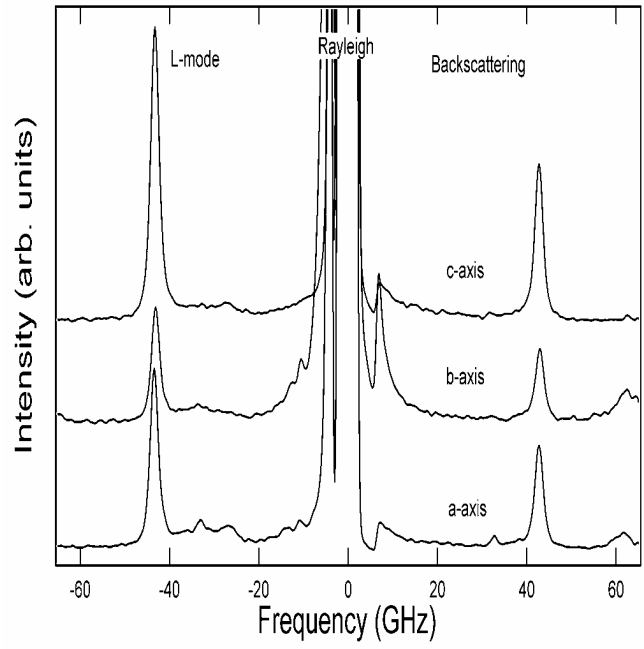

Fig. 4.

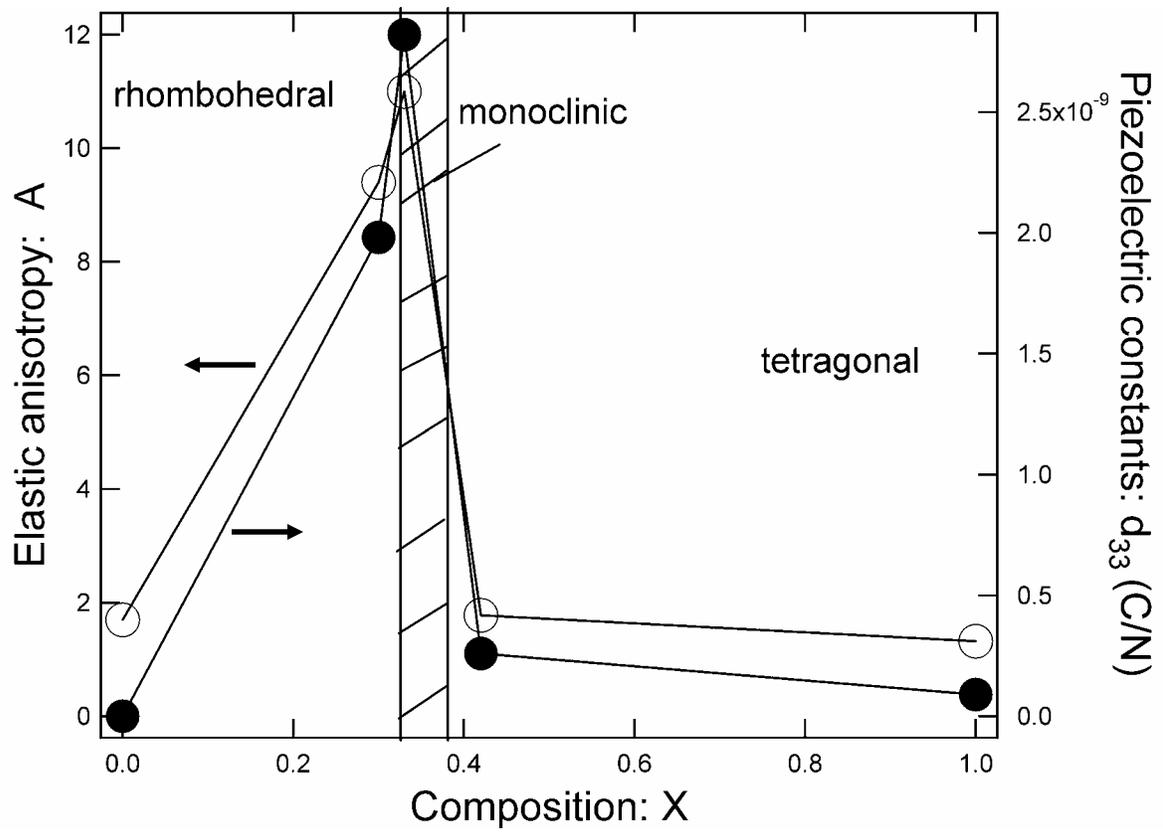

Fig. 5.

**Table I.** Elastic constants for tetragonal $PbTiO_3$ obtained from several different methods. Our MD results are obtained at T = 300 K and zero stress, similar to the experimental results. The DFT results are evaluated at the experiment lattice parameters. A unique solution for the experimental elastic constants could not be derived from the Brillouin scattering data [38]. See text for details on the two methods to extract elastic constants from MD simulations.

| method | $C_{11}$ | $C_{12}$ | $C_{13}$ | $C_{33}$ | $C_{44}$ | $C_{66}$ |
|---|---|---|---|---|---|---|
| DFT [44] | 230 | 96.2 | 65.2 | 41.9 | 46.6 | 98.8 |
| Experiment [38] | 237 | 90 | 70/100 | 60/90 | 69 | 104 |
| Our MD Direct method | 244 ± 13 | 119 ± 14 | 87 ± 9 | 85 ± 12 | 72 ± 7 | 134 ± 3 |
| Our MD σ-ε method | 245 ± 1 | 108 ± 1 | 81 ± 2 | 77 ± 2 | 71 ± 4 | 139 ± 1 |

**Table II.** Material constants for cubic PMN at room temperature obtained from Brillouin.

| | This study | Literature values | | |
|---|---|---|---|---|
| | Fit ($R^2$=0.98) | Ref. 13[a] | Ref. 13 | Ref. 9 |
| Elastic stiffness constants (GPa) | | backscattering | 32° scattering | x-ray |
| $C_{11}^E$ | 156.2 ± 3.4 | 162 | 149 | |
| $C_{12}^E$ | 76.0 ± 3.9 | | | |
| $C_{44}^E$ | 68.5 ± 3.1 | | 67.6 | |
| K(bulk) | 102.7 ± 4.0 | | | 103 |
| A | 1.7 | | | |
| Elastic compliance constants ($10^{-3}$/GPa) | | | | |
| $S_{11}$ | 9.39 | | | |
| $S_{12}$ | -3.07 | | | |
| $S_{44}$ | 14.6 | | 14.3 | |

[a]Brillouin scattering.

**Table III.** Elastic constants obtained from Brillouin compared with MD results for PMN at 300 K. The MD simulations used the random site model to represent the disorder on the B cation in PMN. Elastic constants for five distinct realizations of the random site model are shown.

| Methods | $C_{11}$ | $C_{12}$ | $C_{66}$ | K(bulk) |
|---|---|---|---|---|
| Brillouin | 156.5 | 76 | 68.5 | 103 |
| rnd1 (MD Direct) | 216 ± 15 | 113 ± 8 | 89 ± 5 | 148 |
| rnd1 (MD σ-ε) | 217 ± 3 | 112 ± 2 | 91 ± 5 | 147 |
| rnd2 (MD σ-ε) | 217 ± 5 | 109 ± 3 | 91 ± 6 | 145 |
| rnd3 (MD σ-ε) | 209 ± 5 | 105 ± 2 | 91 ± 2 | 140 |
| rnd4 (MD σ-ε) | 212 ± 1 | 108 ± 2 | 91 ± 3 | 143 |
| rnd5 (MD σ-ε) | 207 ± 5 | 107 ± 2 | 93 ± 2 | 141 |

**Table IV.** Elastic constants for various models of PMN obtained from MD simulations at T= 300K and zero pressure. For models that incorporate disorder (i.e. random site and random) the results are from five different realizations.

| model | $C_{11}$ | $C_{12}$ | $C_{13}$ | $C_{23}$ | $C_{33}$ | $C_{44}$ | $C_{66}$ |
|---|---|---|---|---|---|---|---|
| Random site | 212 ± 4 | 108 ± 2 | - |  | - | - | 91 ± 4 |
| [001]$_{MNN}$ | 177 ± 12 | 120 ± 6 | 83 ± 6 | 92 ± 8 | 219 ± 5 | - | 92 ± 3 |
| [111]$_{MNN}$ | 207 ± 3 | 102 ± 3 | 110 ± 4 | - | 215 ± 5 | 94 ± 1 | 83 ± 1 |
| [001]$_{NCC}$ | 213 ± 3 | 103 ±1 | 109 ± 2 | 114 ± 3 | 219 ± 3 | 90 ± 3 | 61 ± 16 |
| Random | 195 ± 5 | 92 ± 5 | - | - | - | - | 94 ± 4 |